\documentclass[10pt,twocolumn,twoside]{article}
\usepackage{epsfig}
\usepackage{graphicx}
\usepackage{txfonts}
\usepackage{scicite}
\usepackage{fancyheadings}
\begin{document}
\setcounter{page}{1}
\title{Very-High-Energy Gamma Rays from a Distant Quasar:\\How Transparent Is the Universe?}
\author{The MAGIC Collaboration\thanks{
The complete list of authors and their affiliations appears at the
end of this paper.}}
\date{{\it Science} {\bf 320}, 1752 (2008); DOI: 10.1126/science.1157087}
\maketitle

\medskip
{\bf The atmospheric Cherenkov gamma-ray telescope MAGIC, designed
for a low-energy threshold, has detected very-high-energy gamma
rays from a giant flare of the distant Quasi-Stellar Radio Source
(in short: radio quasar) 3C~279, at a distance of more than 5
billion light-years (a redshift of 0.536). No quasar has been
observed previously in very-high-energy gamma radiation, and this
is also the most distant object detected emitting gamma rays above
50 gigaelectron volts. Since high-energy gamma rays may be stopped
by interacting with the diffuse background light in the universe,
the observations by MAGIC imply a low amount for such light,
consistent with that known from galaxy counts.}

\medskip
Ground-based gamma-ray telescopes are sensitive to the Cherenkov
light emitted by the electromagnetic showers that are produced by
gamma rays interacting in the atmosphere. These telescopes have
discovered, since the first detection (in 1989) of gamma rays in
this energy range (from 100~GeV to several TeV), more than 20
blazars, which are thought to be powered by accretion of matter
onto supermassive black holes residing in the centers of galaxies,
and ejecting relativistic jets at small angles to the line of
sight \cite{blandford72}. Most of these objects are of the BL~Lac
type, with weak or no optical emission lines. Quasar 3C~279 shows
optical emission lines that allow a good redshift determination.
Satellite observations with the Energetic Gamma Ray Experiment
Telescope (EGRET) aboard the Compton Gamma Ray Observatory (CGRO)
had measured gamma rays from 3C~279 \cite{hartman} and other
quasars, but only up to the energies of a few GeV, the limit of
the detector's sensitivity. An upper limit for the flux of
very-high-energy (VHE) gamma rays was quoted in \cite{hess08}.

Using MAGIC, the world's largest single-dish gamma-ray telescope
\cite{lorenz} on the Canary island of La Palma (2200 m above sea
level, 28.4$^\circ$N, 17.54$^\circ$W), we detected gamma rays at
energies from 80 to $>$300~GeV, emanating from 3C~279 at a
redshift of 0.536, which corresponds to a light-travel time of 5.3
billion years. No object has been seen before in this range of VHE
gamma-ray energies at such a distance [the highest redshift
previously observed was 0.212 \cite{M1011}], and no quasar has
been previously identified in this range of gamma-ray energies.

The detection of 3C~279 is important, because gamma rays at very
high energies from distant sources are expected to be strongly
attenuated in intergalactic space by the possible interaction with
low-energy photons ($\gamma + \gamma \rightarrow e^+ + e^-$).
These photons [extragalactic background light (EBL) \cite{hauser}]
have been radiated by stars and galaxies in the course of cosmic
history. Their collective spectrum has evolved over time and is a
function of distance. For 3C~279, the range of newly probed EBL
wavelengths lies between 0.2 and 0.8\,$\mu$m
(ultraviolet/optical). Existing instruments that are sensitive
only to higher gamma-ray energies have so far been unable to probe
this domain; by contrast, MAGIC is specifically designed to reach
the lowest-energy threshold among ground-based detectors.

\begin{figure}[ht!]
\includegraphics[width=\linewidth]{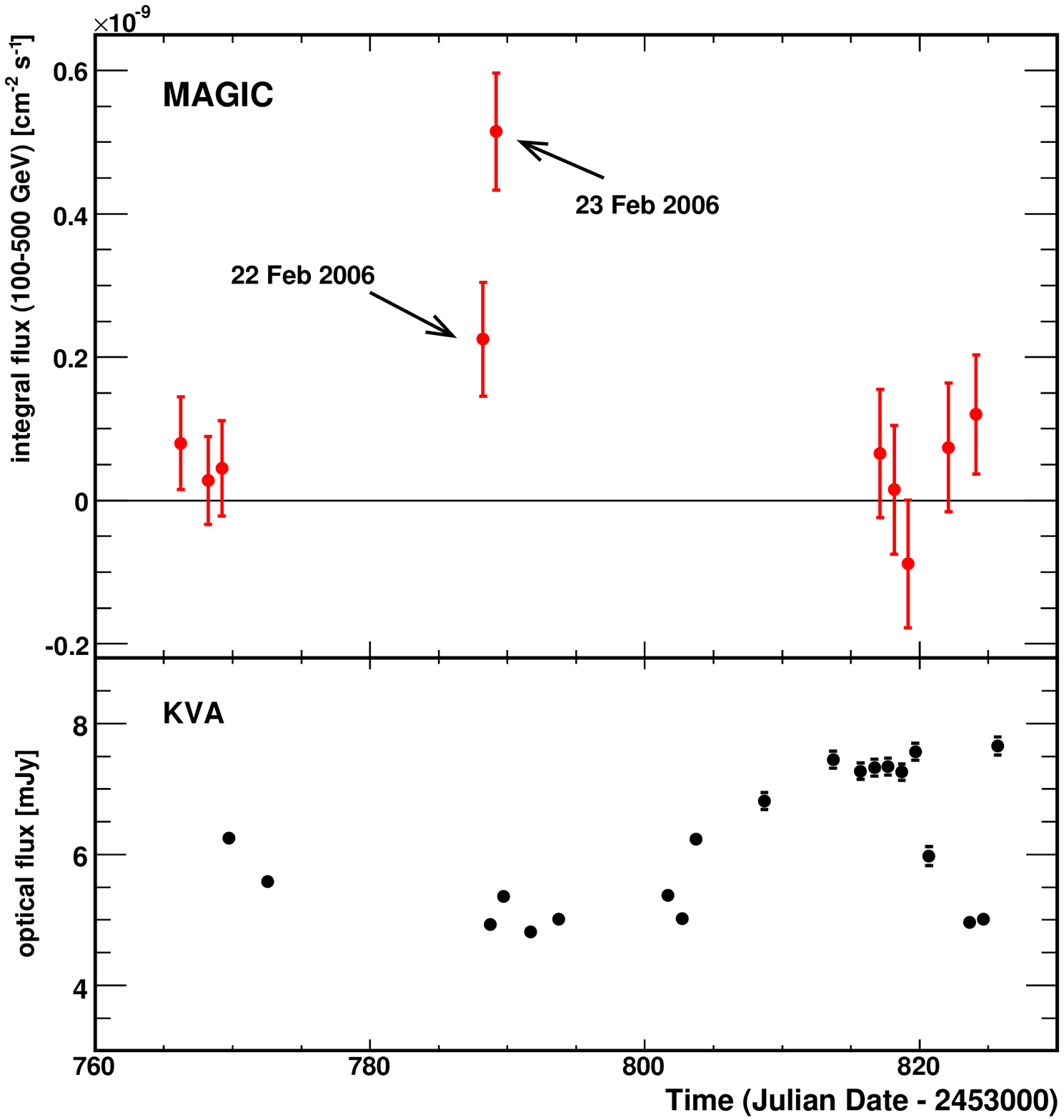}
\textsf{\small {\bf Fig. 1.} Light curves. MAGIC (top) and optical
R-band data (bottom) obtained for 3C~279 from February to March
2006. The long-term baseline for the optical flux is at 3~mJy.}
\end{figure}

\medskip
In observations of 3C~279 over ten nights between late January and
April 2006 (total of 9.7 hours), the gamma-ray source was clearly
detected (at $>$6 SDs) on the night of 23 February, and may also
have been detected the night before (Fig. 1). As determined by the
$\chi^2$ test, the probability that the gamma-ray flux on all 10
nights was zero is $2.3\times10^{-7}$, corresponding to 5.04$\sigma$ in
a Gaussian distribution [see \cite{support}]. Simultaneous optical
R-band observations, by the Tuorla Observatory Blazar Monitoring
Program with the 1.03~m telescope at the Tuorla Observatory,
Finland, and the 35~cm Kungliga Vetenskapsakademien (Royal Swedish
Academy of Sciences) telescope on La Palma, revealed that during
the MAGIC observations, the gamma-ray source was in a generally
high optical state, a factor of 2 above the long-term baseline
flux, but with no indication of short time-scale variability at
visible wavelengths. The observed VHE spectrum (Fig.~2) can be
described by a power law with a differential photon spectral index
of $\alpha=4.1\pm0.7_\mathrm{stat}\pm0.2_\mathrm{syst}$. The
measured integrated flux above 100~GeV on 23~February is $(5.15\pm
0.82_\mathrm{stat}\pm 1.5_\mathrm{syst}) \times 10^{-10}$ photons
cm$^{-2}$ s$^{-1}$.

\begin{figure}
\includegraphics[width=\linewidth]{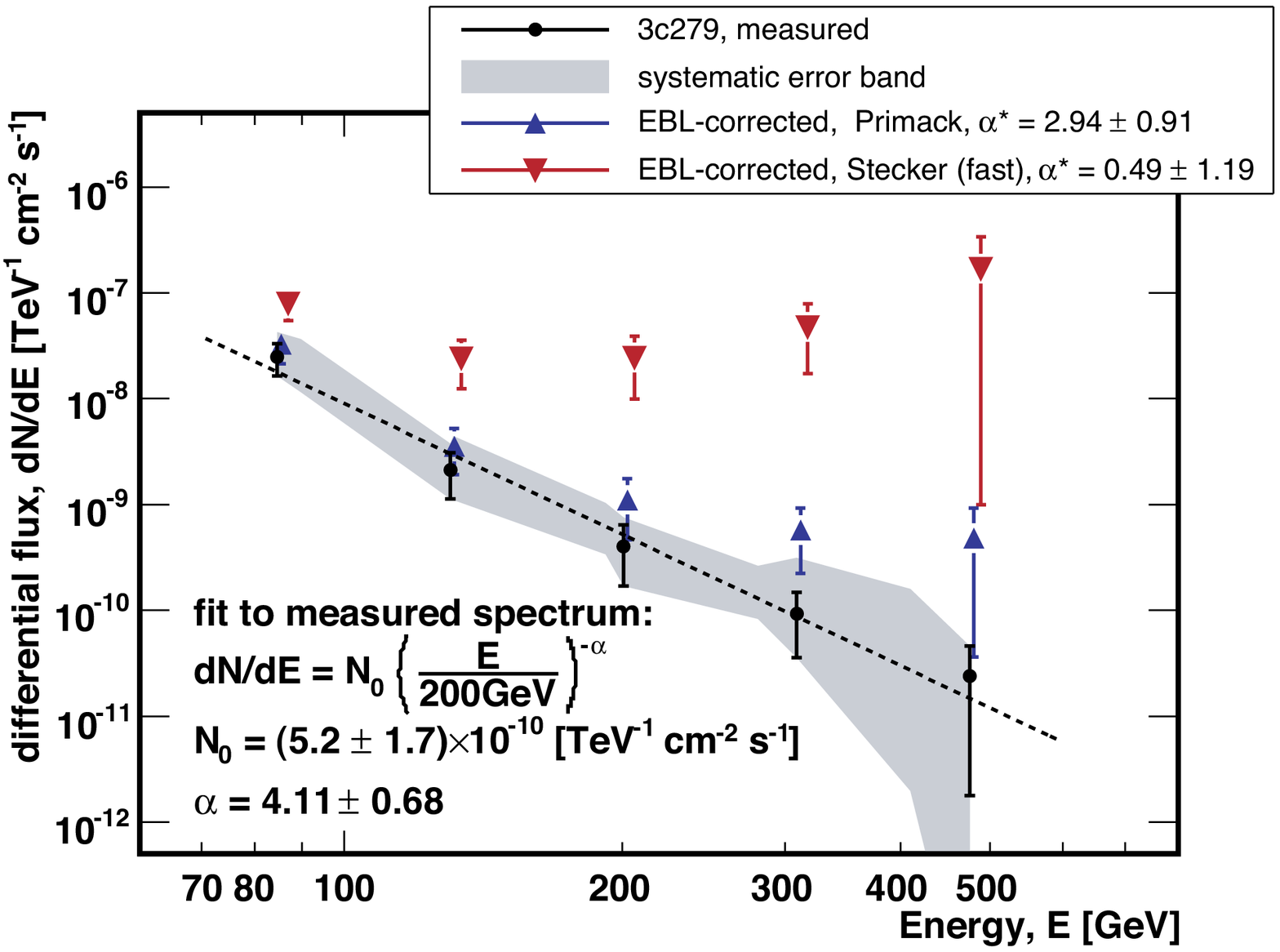}
\textsf{\small {\bf Fig. 2.} Spectrum of 3C~279 measured by MAGIC.
The grey area includes the combined statistical ($1\sigma$) and
systematic errors, and underlines the marginal significance of
detections at high energy. The dotted line shows compatibility of
the measured spectrum with a power law of photon index
$\alpha=4.1$. The blue and red triangles are measurements
corrected on the basis of the two models for the EBL density,
discussed in the text.}
\end{figure}

\medskip
The EBL influences the observed spectrum and flux, resulting in an
exponential decrease with energy and a cutoff in the gamma-ray
spectrum. Several models have been proposed for the EBL
\cite{hauser}. All have limited predictive power for the EBL
density, particularly as a function of time, because many details
of star and galaxy evolution remain uncertain. We illustrate the
uncertainty in the EBL by using two extreme models: a model by
Primack {\it et al.} \cite{primack}, close to the lowest possible
attenuation level consistent with lower EBL limit from galaxy
counts \cite{madau,fazio}; and a ``fast-evolution'' model by
Stecker {\it et al.} \cite{stecker}, corresponding to the highest
attenuation of all the models. We refer to these models as ``low''
and ``high,'' respectively. The measured spectra of 3C~279,
corrected for absorption according to these two models, are shown
in Fig.~2. They represent the range for the possible intrinsic
gamma-ray flux of the source.

A power-law fit to the EBL-corrected points \cite{fn99} results in
an intrinsic photon index of $\alpha^* = 2.9 \pm
0.9_{\mathrm{stat}} \pm 0.5_{\mathrm{syst}}$ (low) and $\alpha^* =
0.5 \pm 1.2_{\mathrm{stat}} \pm 0.5_{\mathrm{syst}} $ (high). The
systematic error is determined by shifting the absolute energy
scale by the estimated energy error of 20\% and recalculating the
intrinsic spectrum. Further discussion of the intrinsic spectrum
and the spectral energy density can be found in \cite{support}.

\medskip
The measured spectrum of 3C~279 permits a test of the transparency
of the universe to gamma rays. The distance at which the flux of
photons of a given energy is attenuated by a factor $e$ (i.e., the
path corresponding to an optical depth $\tau=1$) is called the
gamma-ray horizon and is commonly expressed as a function of the
redshift parameter \cite{fazio_stecker}; we show this
energy/redshift relation in Fig.~3. In the context of Fig.~3, we
make use of a model based on \cite{kneiske} with parameters
adapted to the limits given by \cite{hessnat} and fine-tuned such
that for 3C~279, the intrinsic photon index  is $\alpha^* = 1.5$.
The tuning allows for the statistical and systematic errors (1 SD,
added linearly). Although the intrinsic spectrum emitted by 3C~279
is unknown, $\alpha^* = 1.5$ is the lowest value given for EGRET
sources (not affected by the EBL) and all spectra measured by
gamma-ray telescopes so far \cite{fn100}, so we assume this to be
the hardest acceptable spectrum. The region allowed between the
maximum EBL determined using the above procedure and that from
galaxy counts \cite{primack} is very small.

\begin{figure}
\includegraphics[width=\linewidth]{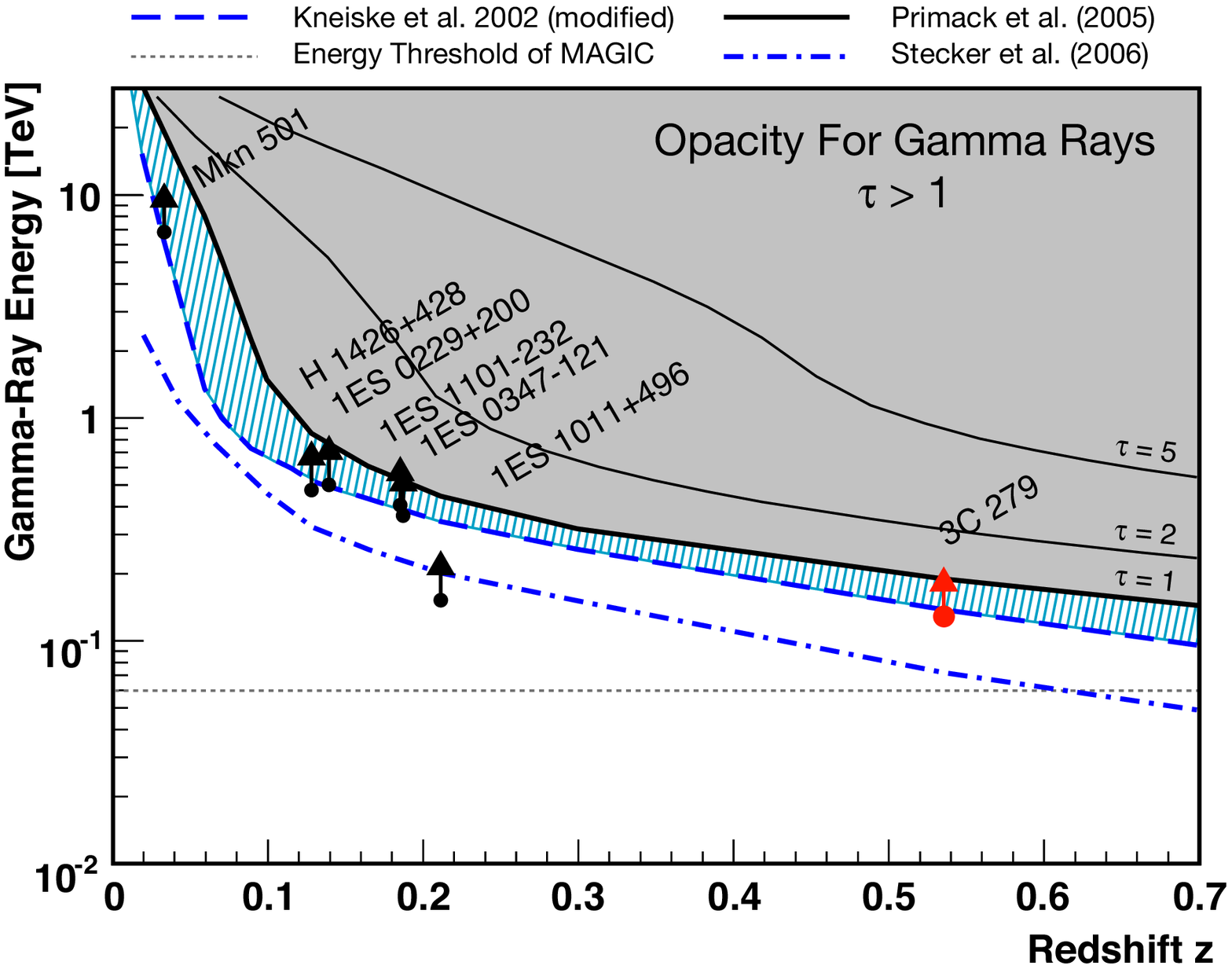}
\textsf{\small {\bf Fig. 3.} The gamma-ray horizon. The redshift
region over which the gamma-ray horizon can be constrained by
observations has been extended up to z=0.536. The prediction range
of EBL models is illustrated by \cite{primack} (thick solid black
line) and \cite{stecker} (dashed-dotted blue line). The tuned
model of \cite{kneiske} (dashed blue line) represents an upper EBL
limit based on our 3C~279 data, obtained on the assumption that
the intrinsic photon index is $\geq 1.5$ (red arrow). Limits
obtained for other sources are shown by black arrows, most of
which lie very close to the model \cite{kneiske}. The narrow blue
band is the region allowed between this model and a maximum
possible transparency (i.e., minimum EBL level) given by
\cite{primack}, which is nearly coincident with galaxy counts. The
gray area indicates an optical $\tau>1$, i.e., the flux of gamma
rays is strongly suppressed. To illustrate the strength of the
attenuation in this area, we also show energies for $\tau=2$ and
$\tau=5$ (thin black lines), again with \cite{primack} as model.}
\end{figure}

The results support, at higher redshift, the conclusion drawn from
earlier measurements \cite{hessnat} that the observations of the
Hubble Space Telescope and Spitzer correctly estimate most of the
light sources in the universe. The derived limits are consistent
with the EBL evolution corresponding to a maximum star-formation
rate at $z \geq 1$, as suggested by \cite{primack} and similar
models.

\medskip
The emission mechanism responsible for the observed VHE radiation
remains uncertain. Leptonic emission models (assuming relativistic
electrons in the jet as source of the gamma rays), generally
successful in describing blazar data [e.g., \cite{maraschi}], can,
with some assumptions, also accommodate the MAGIC spectrum.
Hadronic models [involving relativistic protons,
e.g.\cite{mannbier}] provide a possible alternative. However, a
genuine test of the models can be only obtained with simultaneous
observations at different wavelengths, which are not available for
the observations described here. Future tests of these models
should use observations from sources at all wavelengths from radio
to VHE gamma rays. In the domain of VHE gamma rays, we can expect
important new insights by simultaneous observations with the Large
Area Telescope (LAT), the high-energy gamma-ray instrument on the
Gamma Ray Large Area Space Telescope [GLAST \cite{glast}]. Our
observations of this distant source in VHE gamma rays demonstrate
that a large fraction of the universe is accessible to VHE
astronomy.

{\sf 

\medskip
\noindent{\bf The MAGIC Collaboration:} J.~Albert,$^{1}$
E.~Aliu,$^{2}$ H.~Anderhub,$^{3}$ L.~A.~Antonelli,$^{4}$
P.~Antoranz,$^{5}$ M.~Backes,$^{6}$ C.~Baixeras,$^{7}$
J.~A.~Barrio,$^{5}$ H.~Bartko,$^{8}$ D.~Bastieri,$^{9}$
% ...
\\

\medskip

\noindent{$^1$ Universit\"at W\"urzburg, D-97074 W\"urzburg, Germany.
$^2$ Institut de F\'{\i}sica d'Altes Energies, Edifici Cn., Campus UAB, E-08193 Bellaterra, Spain.
$^3$ ETH Zurich, CH-8093 Switzerland.
$^4$ Istituto Nazionale di Astrofisica (INAF) National Institute for Astrophysics, I-00136 Rome, Italy.
$^5$ Universidad Complutense, E-28040 Madrid, Spain.
$^6$ Technische Universit\"at Dortmund, D-44221 Dortmund, Germany.
$^7$ Universitat Aut\`onoma de Barcelona, E-08193 Bellaterra, Spain.
$^8$ Max-Planck-Institut f\"ur Physik, D-80805 M\"unchen, Germany.
$^9$ Universit\`a di Padova and INFN, I-35131 Padova, Italy.
$^{10}$ University of \L\'od\'z, PL-90236 \L\'od\'z, Poland.
$^{11}$ DESY Deutsches Elektronen-Synchrotron, D-15738 Zeuthen, Germany.
$^{12}$ Universit\`a  di Siena, and INFN Pisa, I-53100 Siena, Italy.
$^{13}$ Universitat de Barcelona Insitut de Ciencias del Cosmos--Institut d'Estudis Especials de Catalunya (IEEC), E-08028 Barcelona, Spain.
$^{14}$ Yerevan Physics Institute, AM-375036 Yerevan, Armenia.
$^{15}$ Instituto de Astrofisica de Canarias, E-38200, La Laguna, Tenerife, Spain.
$^{16}$ Departamento de Astrofisica, Universidad, E-38206 La Laguna, Tenerife, Spain.
$^{17}$ Universit\`a di Udine, and INFN Trieste, I-33100 Udine, Italy.
$^{18}$ Institut de Cienci\`es de l'Espai [IEEC-Consejo Superior de Investigaciones C\'{\i}entificas (CSIC)], E-08193 Bellaterra, Spain.
$^{19}$ Inst. de Astrofisica de Andalucia (CSIC), E-18080 Granada, Spain.
$^{20}$ University of California, Davis, CA-95616-8677, USA.
$^{21}$ Tuorla Observatory, Turku University, FI-21500 Piikki\"o, Finland.
$^{22}$ Institute for Nuclear Research and Nuclear Energy, BG-1784 Sofia, Bulgaria.
$^{23}$ Humboldt-Universit\"at zu Berlin, D-12489 Berlin, Germany.
$^{24}$ INAF/Osservatorio Astronomico and INFN, I-34143 Trieste, Italy.
$^{25}$ Instituc\'{\i}o Catalana de Recerca i Estudis Avancats, E-08010 Barcelona, Spain.
$^{26}$ Universit\`a  di Pisa, and INFN Pisa, I-56126 Pisa, Italy.}
\\

\medskip

\noindent{$^*$To whom correspondence should be addressed.\\E-Mail:
rkbock@gmail.com}

\noindent{$\dag$ Present address: Santa Cruz Institute for
Particle Physics, University of California, Santa Cruz, CA 95064,
USA.}

\noindent{$\ddag$ deceased.}\\

\noindent{27 February 2008; accepted 27 May 2008\\
10.1126/science.1157087 }}

\medskip

\clearpage
\pagestyle{plain}
\noindent{\bf \large Supporting Online Material}\\
www.sciencemag.org/cgi/content/full/320/5884/1752/DC1\\
SOM Text\\
Figs. S1 to S3\\
References
\\
\smallskip

\noindent{\bf Comments to the Introduction}\\
Ground-based gamma ray astronomy, pioneered by the Whipple
Collaboration, has led to a number of important discoveries in
recent years, showing an unexpected wealth of sources in the very
high-energy (VHE) sky above 100~GeV. Among them, some twenty
extragalactic sources were found, all but one are blazars. Blazars
are now known in a broad range of luminosities, from
low-luminosity BL-Lac objects to high-luminosity quasars. They
emit non-thermal radiation from radio waves to gamma rays,
originating presumably in synchrotron emission and inverse-Compton
scattering. Most blazars so far detected with ground-based
telescopes belong to the rare subspecies of BL-Lac objects, that
can be interpreted as a kind of dying quasar running out of fuel,
and thinning out at large cosmological distances
\cite{somcavaliere}. 3C~279 is the first blazar of the quasar type
observed to emit gamma rays in the VHE range. It had been
discovered as a source of lower energy gammas by the EGRET
detector\cite{somhartman}, with a peak flux higher than any other
source seen by EGRET. Subsequently 3C~279 became a much-studied
source with multi-wavelength observation campaigns published
\cite{sommaraschi2,somwehrle,sombach,somcollmar1,somhartman1,somboettcher}.

\medskip
\noindent {\bf MAGIC analysis}\\
The observations were carried out in the ``on/off mode'', which
includes ``on'' events taken with the telescope pointing at the
source, and a similar amount of ``off'' events taken off-axis at a
sky location near the source, with very similar operating
conditions. The ``off'' events are necessary for estimating
reliably the background of hadronic events.

The raw data were calibrated \cite{somgaug} and analyzed with the
standard MAGIC analysis and reconstruction software
\cite{sombretz}. Data runs with anomalous trigger rates due to bad
observation conditions were rejected. The remaining ``on'' data
correspond to 9.7 hours, with 4.9 hours of ``off'' data. The
standard analysis reduces the image to a single cluster of pixels
by removing image noise and tracks far from the main shower, and
calculates image parameters from the cleaned image
\cite{somhillas}. These are used by a multi-variate method to
discriminate gamma rays from background hadrons; the method is
called ``Random Forest'' \cite{somrf} and is a multi-tree
classifier algorithm which calculates, from a combination of image
parameters, a separation variable called hadronness.

All image parameters were checked for consistency between ``on''
and ``off'' data. The excess events were obtained by subtracting
suitably normalized ``off'' data from the ``on'' data, as shown in
Fig.~S1. The variable Alpha is one of the image parameters. It
describes the direction of the main alignment axis of the pixels
in the  image, a zero value indicating a shower seemingly coming
(in the camera plane) from the known position of the source. Alpha
was not included in the multivariate gamma-hadron separation. Cuts
in hadronness and Alpha were optimized using data samples from
observations of the Crab Nebula at comparable zenith angles. For
the detection (and the lightcurve) we require hadronness$<$0.12
and Alpha$<$12$^\circ$, resulting in an efficiency of 40\% for
gammas. For these cuts and with a minimal Size of 100 (Size is the
sum of signals, in photo-electrons), we obtain a peak in the
distribution of estimated energy at 170~GeV.

The significance of the signal of 23 February was calculated
according to equation 17 in \cite{somlima}, resulting in
6.15$\sigma$, taking the difference between a smoothing parabola
fit to normalized ``off'' data and the ``on'' data. The
probability that we deal with a statistical fluctuation in ten
independent observations (trial factor corresponding to 10 nights
of observation) is $3.87 \times 10^{-9}$, which corresponds to a
significance of 5.77$\sigma$. If ``off'' and ``on'' data are
compared bin by bin without smoothing, the significance is
5.5$\sigma$ for 23 February, and 5.1$\sigma$ after correcting for
trials. We also applied a $\chi^2$ test to the results of ten
nights, giving 50.35 for 10 degrees of freedom, or a probability
of $2.3x10^{-7}$, which in a Gaussian distribution corresponds to
5.04$\sigma$. The long-term detector stability has been studied by
monitoring the Crab Nebula over a large range of zenith angles,
and for several years. The observed integrated flux above 100~GeV
has shown no evidence of any variations beyond what is expected
from statistics. We can thus conclude that the observed signal is
indeed an isolated flare from 3C~279.
\begin{figure}[h!]
\includegraphics[width=\linewidth]{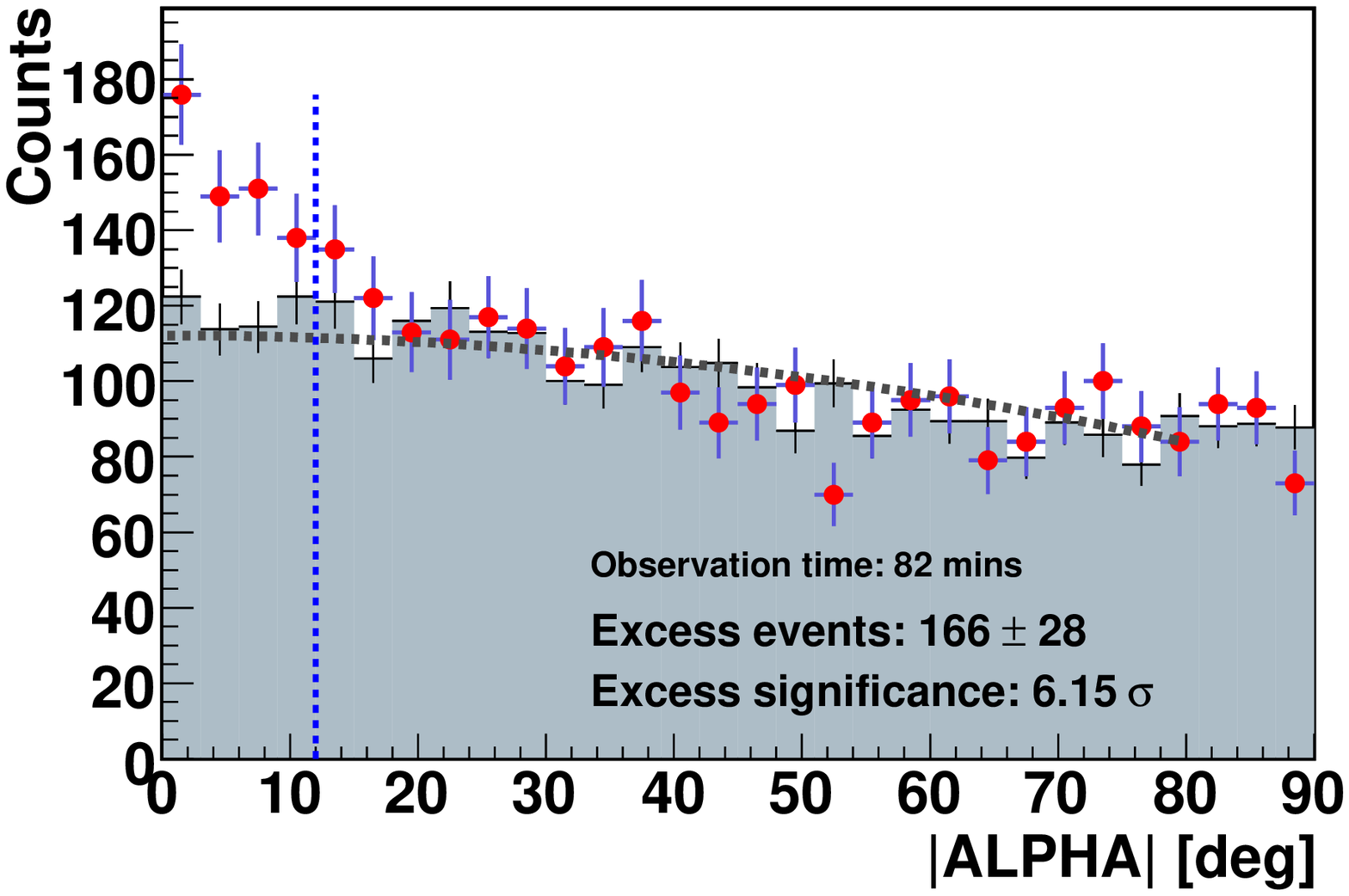}
\textsf{\small {\bf Fig. S1.} Histogram of the image parameter
Alpha. The image parameter Alpha obtained for 3C~279 in the night
of 23 February 2006, for ``on'' (red dots) and ``off'' (crosses
and filled area) pointings, respectively. The dotted line is a
simple parabolic fit (without linear term) to the ``off'' events
between 0$^\circ$ and 80$^\circ$, serving to smoothen the
background for better extrapolation towards Alpha=0; the excess
events are obtained with respect to this line.}
\end{figure}

The reconstruction of gamma ray energy uses an approximation
derived from image parameters. It is obtained by the the Random
Forest  method using Monte Carlo gamma samples. For the
reconstruction of the energy spectrum, we used a low-energy
analysis, making also use of the timing properties of the air
showers with a lower energy threshold ($\approx$110~GeV). The cuts
for obtaining the spectrum were slightly loosened, resulting in a
gamma efficiency of 50\%. The obtained spectrum was unfolded for
detector effects using \cite{somunfold}. The spectrum was
cross-checked with two alternative methods (one of them is the
standard analysis and a second one also uses of the timing
information). All three spectra agree well, as shown in the figure
(main paper) as grey area, the envelope shape for 1$\sigma$ errors
of the spectra. The predicted energy resolution is $\approx$23\%,
reasonably constant above 150~GeV. Systematic errors of our
detector (from atmospheric density and transmission uncertainties,
the photon detection efficiency, non-linearity in the signal
chain, limits in Monte Carlo simulations, choices in analysis
cuts, etc.) are discussed in detail in \cite{somcrab}.
Uncertainties result in a possible systematic energy shift of up
to $\pm$20\%, which translates into a shift of $\mp$25\% in the
limit set for the gamma ray horizon (energy for which the optical
depth~$\tau$~=~1). The uncertainty in the spectral slope
coefficient is $\pm$0.2.

\medskip
\noindent {\bf Attenuation of gamma rays from distant sources and EBL models}\\
Absorption of high-energy gammas by the extragalactic background
light (EBL) depends on the column density of low-energy photons
traversed by the gamma rays.  For gamma rays of observed energy
$E_\gamma$, the maximal cross section for pair production at
redshift $z$ occurs at wavelength $\lambda=1.24\,\mu{\rm
m}(E_\gamma/\rm 1\,\mathrm{TeV})(1+z)^2$. Folded with the shape of
the cross section, MAGIC's energy range specifically probes (with
3C~279) background wavelengths of $0.2~\mu\rm m-2~\mu m$. This
highlights the importance of the ultraviolet-to-near-infrared
wavelength range in modifying the spectra of sources at gigaparsec
distances \cite{somhessnat}. Knowledge of the amount of EBL along
the line of sight to gamma ray sources is based on bolometric
measurements and galaxy counts. Bolometric measurements are
generally hampered by foreground emission in our own Galaxy
\cite{sommattila}. Since the light emitted from galaxies changes
with time following the star formation history \cite{sommadau} and
with the ageing stellar populations, the EBL has changed over
time, too. An in-depth review can be found in \cite{somhauser}.

Various models for the EBL have been published. In
\cite{somkneiske}, several parameterized models using different
assumptions for the evolving EBL are introduced, taking into
account magnetic fields, gas and dust, the relative abundance of
elements heavier than Helium, and the star formation rate. A range
of parameters is proposed such that the results agree with what is
known from deep galaxy surveys and infrared background
measurements available in 2002. In a recent publication
\cite{somstecker}, two different backward-evolution EBL models
were used to calculate the intergalactic photon density from $z=0$
to $z=6$ (called ``Base'' and `´Fast'' evolution). In this report
we use the ``Fast'' evolution model as predicting maximally {\it
high} absorption. In a different publication \cite{somprimack},  a
semi-analytical forward-evolution approach was used to obtain the
evolving EBL density. This model predicts a rather {\it low} EBL
level at $z=0$, just on top of the galaxy counts obtained from the
Hubble Space Telescope \cite{sommadau}. In fact, the model
\cite{somprimack} predicts the EBL level at wavelengths above
$5~\mu\mathrm{m}$ slightly below the galaxy counts obtained from
the Spitzer \cite{somfazio, sompapovich, somdole} data and the
ISOCAM results \cite{sommetcalfe, somelbaz}. The model
\cite{somprimack} also lies below the FIRAS measurements in the
far-infrared, an undershoot attributed to the abundance of
Ultra-Luminous Infrared Red Galaxies (ULIRG), only recently
discovered and not taken into account in the model. However, the
EBL domain above $3~\mu\mathrm{m}$ does not influence our result,
given that the highest energy measured by MAGIC in the 3C~279
spectrum is below 500~GeV; EBL photons with $\lambda >
2.5~\mu\mathrm{m}$ do not reach the energy necessary for pair
production. We have, therefore, chosen the model \cite{somprimack}
as the {\it low} limit for the EBL level in the frequency range
relevant for the MAGIC 3C~279 measurement.

In the discussion of the gamma ray horizon in the main paper, we
have used a model tuned to give for 3C~279 the intrinsic photon
index  $\alpha^* = 1.5$. We call this ad-hoc EBL model maxEBL and
give more details in the following. It is based on
\cite{somkneiske}, with modified parameters. The "best-fit" model,
defined in \cite{somkneiske04}, is changed by reducing the optical
and infrared star formation rate, by adding a "warm dust"
component and by taking into account a high fraction of UV
emission, which fits the data derived from AGN absorption lines, a
method called proximity effect [e.g., \cite{somscott}].
Additionally, the dust parameter for the optical galaxy component
is set to a very low value of E(B-V) = 0.01, to account for the
maximum possible optical flux.
In detail, the parameters used [definition in \cite{somkneiske}] are the following:\\
$\mathrm{SFR}_\mathrm{OPT}: \alpha=3.5, \beta=-1.2, z_p=1.2, \dot{\rho_\ast}(z_p)=0.08,$\\
$\mathrm{SFR}_\mathrm{LIG}: \alpha=4.5, \beta=0.0, z_p=1.0, \dot{\rho_\ast}(z_p)=0.09,$\\
and further $f_\mathrm{esc}=4, c_2 = 10^{-23.4},
E(B-V)_\mathrm{OPT} = 0.01.$ Contrary to the "best-fit" EBL, the
maxEBL model is above the emissivity data at wavelengths between
0.16 -- 1.0 $\mu$m [\cite{somkneiske} and references therein].

All EBL models mentioned contain the evolution of the radiation
contributing to the EBL at different redshifts, and use the
Lambda-CDM Universe with $\Omega_{\lambda} = 0.7$ and
$\Omega_{\mbox{m}} = 0.3$, confirmed by the latest results from
the Wilkinson Microwave Anisotropy Probe (WMAP) \cite{somwmap}.
The models are graphically summarized in Fig.~S2, along with a
number of measurements.

\begin{figure}[h!]
\includegraphics[width=\linewidth]{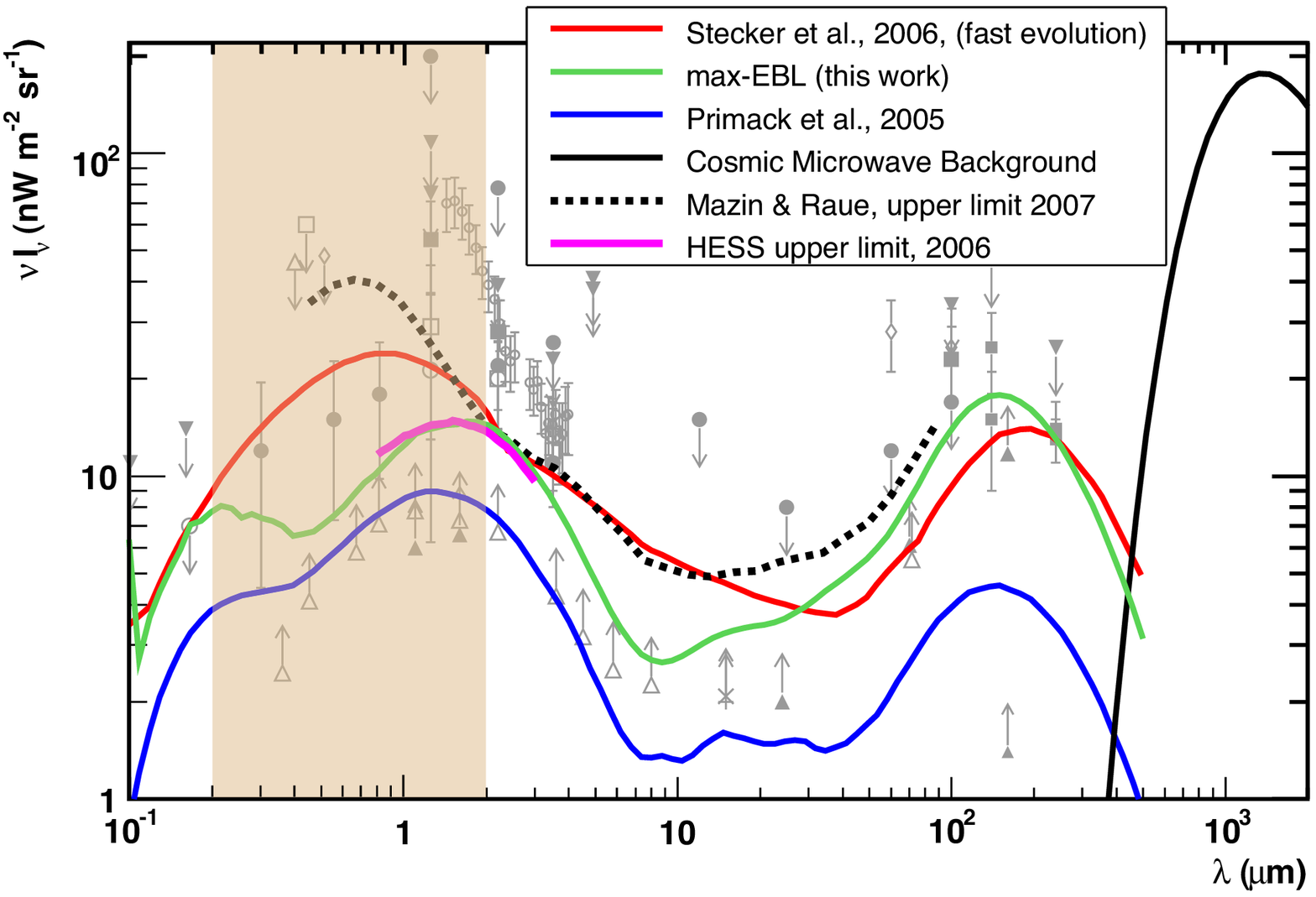}
\textsf{\small {\bf Fig. S2.} EBL Models. Some of the models for
EBL, for z=0, and measurements at various wavelengths. The green
line is based on \cite{somkneiske}, with parameters adapted to the
results from the latest galaxy counts. The dotted line is from
\cite{sommazin}. The shaded vertical band indicates the range of
frequencies corresponding to the MAGIC observations.}
\end{figure}

\pagestyle{fancy}

Comparing the effect of different EBL models on our measured spectrum
we add the following comments:
\begin{itemize}
 \item The {\it high} model \cite{somstecker} leads to an
   intrinsic spectrum of 3C~279 with a (fitted) index of $\alpha^*=0.5\pm1.2$
   (see main paper), a rise in the VHE gamma ray
   spectrum very difficult to reconcile with extrapolation from EGRET data
   and with general constraints in the spectral energy distribution.
 \item The maxEBL model has been derived to be in agreement with an intrinsic blazar
   spectral index of $\alpha^*=1.5$, and can be considered as an upper
   limit on the EBL density respecting the constraints from the measured
   VHE blazar spectra \cite{somhessnat,sommazin}.
   Note that the values for the optical and
   infrared star formation rate (SFR) and for the optical dust parameter are close to their
   lower limits, and no AGN contribution has been taken into account.
 \item The {\it low} EBL derived from \cite{somprimack} apparently gives an
   acceptable result,
   resulting in a most probable intrinsic power law with a spectral index
   $\alpha^*=2.9\pm0.9$ (see main paper).
\end{itemize}
Some of the models discussed above seem less likely in the light
of the present measurements, requiring the sources to exhibit a
redshift-dependent hardening of their spectra. We should also note
that the EBL density may be higher than derived here, if VHE
photons are speculated to undergo interactions beyond standard
physics [e.g., \cite{somdeangel}].

\medskip
\noindent {\bf Spectral Energy Distribution}\\
Blazars represent the most violent variant of the AGN phenomenon,
related to the activity of supermassive ($M=10^8-10^9$ solar
masses) black holes harbored in the center of galaxies. Their
emission, extending from radio wavelengths to high-energy gamma
rays, is produced within a jet of plasma ejected at relativistic
speeds and pointing close to the Earth.

The spectral energy distribution (SED) of blazars is characterized
by two broad bumps, the first one peaking between the infrared and
the X-ray band, the second one in the gamma ray domain. The first
peak traces the synchrotron emission of relativistic electrons
spiraling along the lines of the magnetic field in the jets; the
high-energy peak is presumably due to the inverse Compton
scattering between these relativistic electrons and low-energy
photons. The latter can be the synchrotron photons themselves
[Synchrotron-Self-Compton (SSC) mechanism \cite{sommaraschi}] or
ambient radiation [External-Compton (EC) mechanism
\cite{somdermer,somsikora}]. Ambient radiation can enter the jet
either from an accretion disk or from surrounding gas clouds.
These models ultimately rely on electrons (or pairs) accelerated
within the flow and thus are called {\it leptonic models}.

Other possibilities ({\it hadronic models}) involve gamma ray
emission due to accelerated protons and ions. Such emission can
result from pion production in hadronic interactions, pair
production, or be due to synchrotron radiation [e.g.,
\cite{sommannheim,sommannbier,somahar2000,sommucke}], and includes
the associated production of neutrinos and ultrahigh-energy cosmic
rays \cite{sommannstan}. Although the energy losses of protons and
ions are suppressed by their large mass and the low matter density
in the acceleration zone, they become sizeable at ultrahigh
energies where photo-production of pions can take place in
collisions with low-energy synchrotron photons.

Until now the majority of known extragalactic sources of VHE
photons belongs to the class of BL-Lac objects, whose spectra show
a weak signature of thermal radiation from outside the jet and
hence favor the SSC scenario. In quasars [mostly Flat Spectrum
Radio Quasars (FSRQs)], such as 3C~279 the situation with thermal
radiation is different (the {\it blue bump} in the spectrum), and
the EC process is likely to dominate if the gamma ray emission
zone is close to the origin of the jet. The validity of the EC
mechanism is not undisputed, however, since the gamma ray emission
zone could well be far away from the central source of thermal
radiation [e.g., \cite{somlindfors1,somlindfors2}].

Hadronic emission models require much larger magnetic field
strengths than leptonic models, in order to confine the more
massive and energetic particles to the jet.  They generically
produce harder VHE spectra than leptonic models. In low-peaked
blazars at high redshifts such as FSRQs, however, the spectrum is
steepened due to attenuation by the EBL, rendering differences in
the spectra difficult to detect. 3C~279 is among the nearest
FSRQs, and the energy threshold of MAGIC is low enough to permit
observing this source.

The EGRET gamma ray telescope (operating in the past decade and
sensitive between 20~MeV and few GeV) observed at several
occasions bright and hard (photon index around 2) emission from
3C~279. In particular, during an extensive campaign held in 1996
the source showed a large (a factor 10 in amplitude) and short
($\sim$ 2-3 days) gamma ray flare \cite{somwehrle} with episodes
of relatively fast variability, including an increase by a factor
of 2.6 in $\sim$8~hours. Naively extrapolating this behavior to
the MAGIC band one could expect bright and short episodes also in
this band. However, in the standard leptonic framework the
detection of photons with energy above 100~GeV emitted by 3C~279
is somewhat surprising. Indeed, a careful consideration of the
conditions expected in the source raises several problems, related
to the high energy required for the electrons to emit photons with
such high energy and the strong opacity to gamma rays expected
inside these sources. The detection of strong VHE emission from
3C~279 thus is stretching leptonic models to or beyond their
limits.

\medskip
\noindent {\bf Emission model}\\
The simplest EC model (``one-zone'') assumes that the bulk of the
emission is produced within a single region. In order to properly
model the emission, good coverage of the entire SED is required,
since only with known position and luminosity of both peaks is it
possible to fully constrain the physical parameters [e.g.,
\cite{somtavecchio}]. In the case of the 3C~279 flare, we only
have the simultaneous measurement in the optical R--band with the
VHE spectrum derived by MAGIC. With this input we tried to
reproduce the observed VHE spectrum assuming a simple one-zone
model [see \cite{sommaraschi1} for a full description].

The MAGIC observations contribute data to the important energy
range above $10^{25}$Hz, poorly studied thus far because lying
above the range of satellite-borne instruments and below that of
most ground-based Cherenkov telescopes. The SED in Fig.~S3 shows
the MAGIC points corrected according to the two extreme EBL models
discussed in the paper, together with measurements at lower
frequencies from different epochs. In the same figure we show the
generic one-zone model described here. The model uses the
following assumptions and parameters: The source is spherical with
radius $R=5\times 10^{16}$ cm, and it is in motion with a bulk
Lorentz factor $\Gamma=20$ at an angle $\theta =\Gamma^{-1}=2.9$
deg with respect to the line of sight. The magnetic field has an
assumed intensity $B=0.15$~G. The particles, with a total density
$n=2\times 10^4$ cm$^{-3}$, follow an energy distribution
represented by a broken power law extending from $\gamma _1=1$ to
$\gamma _2=3\times 10^5$, with indices $n_1=2$ and $n_2=3.7$ below
and above the break at $\gamma _p=2.5\times 10^3$. We model the
external radiation field assuming that a fraction $\tau =0.005$ of
the disk emission, a black body with luminosity $6\times 10^{45}$
erg/s, is diluted into the Broad Line Region, a sphere with radius
$4\times 10^{17}$ cm.

\begin{figure}[h!]
\includegraphics[width=\linewidth]{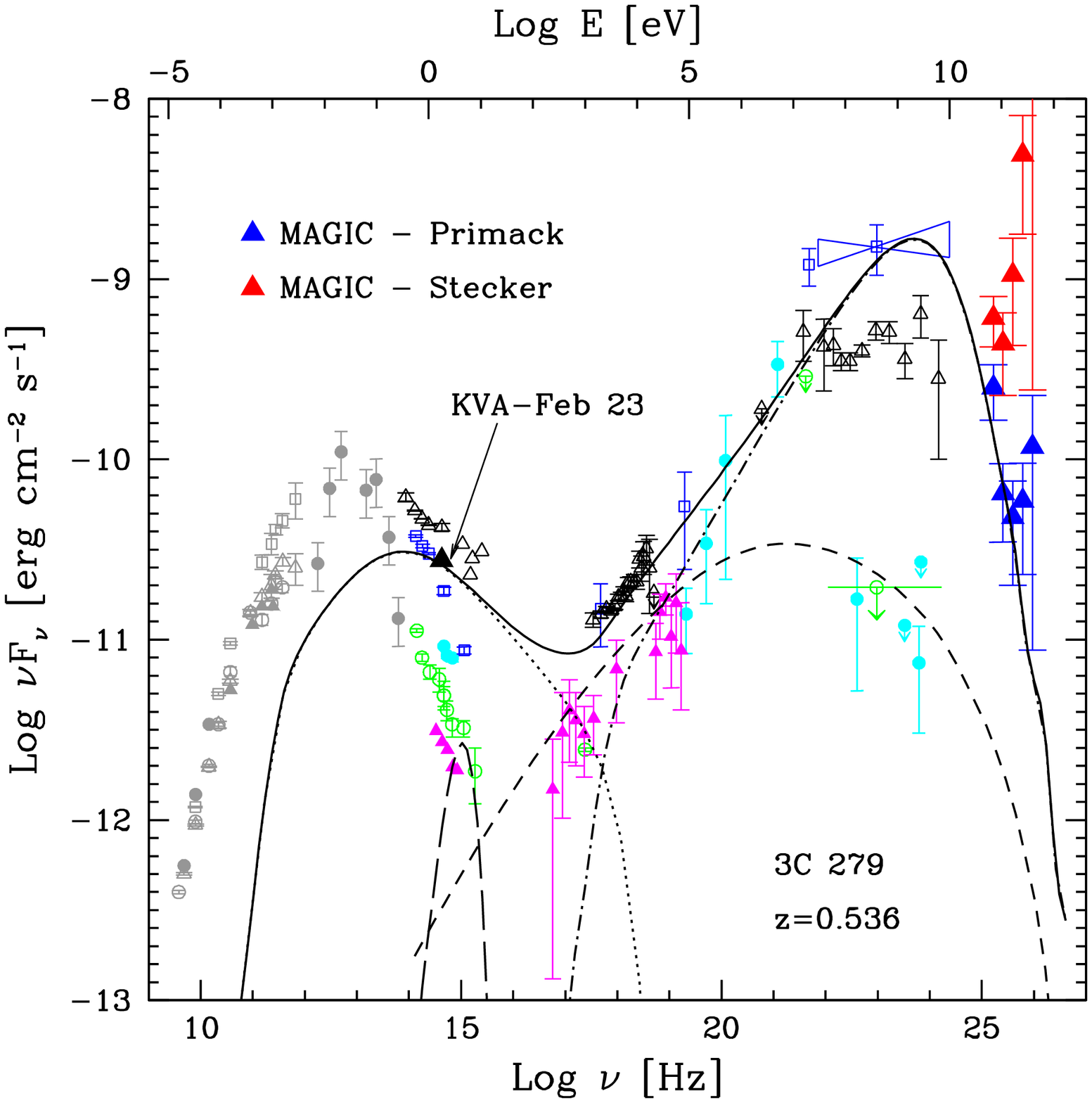}
\textsf{\small {\bf Fig. S3.} Spectral energy distribution for
3C~279. Observations at different source intensities, over the
years from 1991 to 2003, from \cite{somballo} and
\cite{somcollmar}, with MAGIC points (2006) at the high-energy end
(right). The MAGIC points are corrected according to the two
extreme EBL models of \cite{somprimack} and \cite{somstecker}
(blue and red triangles, respectively); the former clearly gives a
more probable result with a spectral index of
$\alpha^*=2.9\pm0.9$. An emission model and its individual
components are also shown: the solid line represents the total
model emission, fitted only to the (blue) MAGIC points and the
(black) triangle from the KVA telescope. The individual components
of the emission model are synchrotron radiation (dotted line),
disk emission (long-dashed), synchrotron-self-Compton
(short-dashed), and external Compton (dot-dashed).}
\end{figure}

\medskip
\noindent {\bf Inverse-Compton peak}\\
Accounting for attenuation of the gamma rays due to pair
production in collisions with photons from the integrated light of
galaxies, brings the emitted spectrum into agreement with models
assuming inverse-Compton scattering of external photons as the
dominant radiation process. The emitted spectrum also appears to
be very similar to the spectra of many high-peaked BL-Lac objects,
although their corresponding synchrotron peak is at much higher
energies. In the framework of inverse-Compton scattering models
this implies that external photon fields may act as scattering
target, although seemingly not to the extent to create internal
opacity due to pair production in photon-photon collisions
\cite{somreimer}. The inverse-Compton models predict a steep
decline for two reasons: for one, the spectrum of the highest
energy electrons, judging from the synchrotron spectrum, tends to
be steep; and second, VHE electrons scatter with ambient photons
in the Klein-Nishina regime, reducing the scattering efficiency.
An EBL larger than the integrated light of known galaxies would
imply that the generic leptonic emission models must undergo a
major revision, perhaps involving proton synchrotron and cascade
emission. The latter would shed new light on the correlation
between active galactic nuclei and the directions of the highest
energy cosmic rays \cite{somauger}.

{\sf 
}


\begin{thebibliography}{10}
\bibitem{fn99} We approximate the intrinsic energy spectrum by $dN/dE \propto$ E$^{-\alpha^*}$, where $\alpha^*$ is the intrinsic photon spectral index.
\bibitem{fn100} According to recent simulations \cite{stecker07}, photon indices of $<$.5 can not be entirely excluded.
\end{thebibliography}
\end{document}